\documentclass[11pt]{article}

\usepackage{graphicx}
\usepackage{amssymb}
\usepackage{amsmath}
\usepackage{upgreek}
\usepackage{multirow}
\usepackage[latin1]{inputenc}
\usepackage[labelfont=bf,small,sf,labelsep=quad,textfont=sf]{caption}
\usepackage{color}
\usepackage[breaklinks=true,colorlinks=true,linkcolor=blue,citecolor=blue,backref=none,pagebackref=false]{hyperref}
\usepackage{setspace}
\usepackage{epstopdf}

\pdfoutput=1
\usepackage{geometry}
\usepackage[auth-sc-lg,affil-sl]{authblk}

\usepackage{setspace}
\newcommand{\keywords}[1]{\begin{center}\textbf{Keywords:} #1\end{center}}

\newcommand{\rbar}{\overline{r}}
\usepackage{mychicago}
\usepackage{genetics_manu_style}

\begin{document}

\title{Stochastic tunneling and metastable states\\during the somatic evolution of cancer}

\author[1]{Peter Ashcroft \thanks{peter.ashcroft@postgrad.manchester.ac.uk}}
\affil[1]{Theoretical Physics, School of Physics and Astronomy, The University of Manchester, Manchester M13 9PL, United Kingdom}

\author[2]{Franziska Michor \thanks{michor@jimmy.harvard.edu}}
\affil[2]{Department of Biostatistics and Computational Biology, Dana-Farber Cancer Institute, and Department of Biostatistics, Harvard School of Public Health, Boston, MA 02215, USA}

\author[1]{Tobias Galla \thanks{tobias.galla@manchester.ac.uk} \thanks{Corresponding author}}

\maketitle

%% ABSTRACT ENVIRONMENT
\begin{abstract}
Tumors initiate when a population of proliferating cells accumulates a certain number and type of genetic and/or epigenetic alterations. The population dynamics of such sequential acquisition of (epi)genetic alterations has been the topic of much investigation. The phenomenon of stochastic tunneling, where an intermediate mutant in a sequence does not reach fixation in a population before generating a double mutant, has been studied using a variety of computational and mathematical methods. However, the field still lacks a comprehensive analytical description since theoretical predictions of fixation times are only available for cases in which the second mutant is advantageous. Here, we study stochastic tunneling in a Moran model. Analyzing the deterministic dynamics of large populations we systematically identify the parameter regimes captured by existing approaches. Our analysis also reveals fitness landscapes and mutation rates for which finite populations are found in long-lived metastable states. These are landscapes in which the final mutant is not the most advantageous in the sequence, and resulting metastable states are a consequence of a mutation-selection balance. The escape from these states is driven by intrinsic noise, and their location affects the probability of tunneling. Existing methods no longer apply. In these regimes it is the escape from the metastable states that is the key bottleneck; fixation is no longer limited by the emergence of a successful mutant lineage. We used the so-called Wentzel-Kramers-Brillouin method to compute fixation times in these parameter regimes, successfully validated by stochastic simulations. Our work fills a gap left by previous approaches and provides a more comprehensive description of the acquisition of multiple mutations in populations of somatic cells.
\end{abstract}

\newpage

\onehalfspacing

\section{Introduction}
Understanding the dynamics of an evolving population structure has long been the goal of population genetics. Several authors have constructed probabilistic models to study allele frequency distributions in populations subject to mutation, selection, and genetic drift \cite{fisher1930genetical,wright1931evolution,moran1962statistical}. The mathematical analysis of these models leads to an improved understanding of the underlying system, and has been crucial for the interpretation of the laws of evolution. This is most evident in the quantitative analysis of cancer, which has seen numerous studies throughout the 20th century that addressed the kinetics of cancer initiation and progression \cite{nordling,armitagedoll,fisher,knudson1971mutation,moolgavkar}. Due to these and other studies (see \citeN{weinberg} for a review), we now know that human cancer initiates when cells within a proliferating tissue accumulate a certain number and type of genetic and/or epigenetic alterations. These alterations can be point mutations, amplification and deletion of genomic material, structural changes such as translocations, loss or gain of DNA methylation and histone modifications, and others \cite{weinberg}.

The dynamics of mutation acquisition is governed by evolutionary parameters such as the rate at which alterations arise, the selection effect that these alterations confer to cells, and the size of the population of cells that proliferate within a tissue. Much effort has been devoted to model these processes mathematically and computationally, and to analyze the rates at which mutations arise within pre-cancerous tissues \cite{moolgavkarluebeck,nunney,gatenbyvincent,michor2004dynamics,haeno2009}.
In particular, several investigators have studied the dynamics of two mutations arising sequentially in a population of a fixed finite number of cells. This scenario describes the inactivation of a tumor-suppressor gene (TSG) which directly regulates the growth and differentiation pathways of the cells \cite{weinberg}. This may or may not lead directly to cancer.
Cells in which the TSG is inactivated can take a variety of fitness values. For instance embryonic retina cells with an inactivated RB1 gene can proliferate uncontrollably and create retinoblastomas \cite{knudson1971mutation}. By definition these cells have a higher fitness than the wild-type cells. Alternatively, if chromosomal instability (CIN) is taken into account, cells with deactivated TSG can have a lower fitness than the wildtype \cite{michor2005can}. Empirical evidence for the exact fitness (dis)advantage conferred to cells as a result of accumulating mutations is in general difficult to obtain, since {\em in vitro} growth assays of non-transformed cells are challenging. For this reason and in order to provide general methods, the modeling literature has addressed a range of fitness values for single- and double-mutant cells (e.g. \citeN{michor2004dynamics}).

\begin{figure}
\begin{center}
\includegraphics*[width=0.8\textwidth]{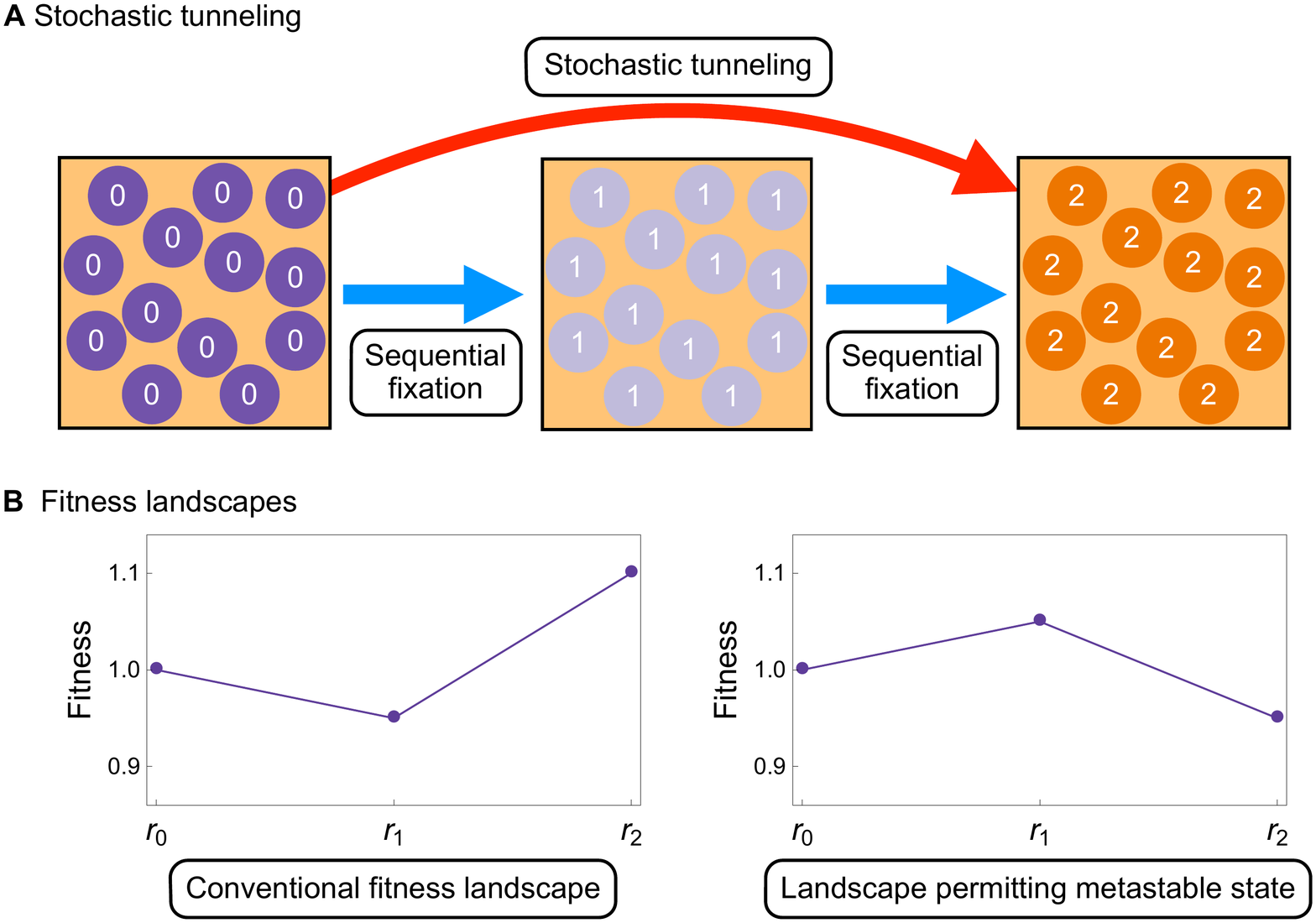}
\end{center}
\caption{\textbf{Stochastic tunneling and fitness landscape examples}
\textbf{A} Schematic of stochastic tunneling.
The population can reach the all-2 state via two routes. The first is the sequential fixation route in which the first mutation takes over the population, and where this is then followed by the second mutation. The second route does not visit the all-1 state. This is the stochastic tunneling route. The arrows are related to the change of state only, and imply nothing about the fitness of these states.
\textbf{B} Examples of fitness landscapes. The first example is the conventional landscape, with $r_1\le r_0$ and $r_2>\max(r_0,r_1)$. The second example has $r_1>r_2$, and as a result this landscape permits a metastable state (a fixed point in the deterministic dynamics as discussed below).}
\label{fig:fig1}
\end{figure}

Subsequent modeling work on mutation acquisition \cite{komarova2003mutation,iwasa2004stochastic,nowak2004evolutionary,proulx2011rate,haeno2013stochastic} has revealed a more detailed picture; a homogeneous population harboring no mutations can move to a homogeneous state in which all cells carry two mutations without ever visiting a homogeneous state in which all cells harbor just one mutation. This phenomenon is referred to as `stochastic tunneling' and represents an additional route to the homogeneous state with two mutations; the sequential route is still available to the system, but it becomes less likely in certain parameter regimes. In this context the term `tunneling' refers only to overlapping transitions between the homogeneous states, it does not imply a statement about the structure of the underlying fitness landscape. The process we refer to as `tunneling' is not limited to valley-crossing scenarios. Fig.~\ref{fig:fig1}A provides a schematic illustration of the tunneling process. 

As with much of the existing literature on the stochastic tunneling, our work is not just limited to the case of cancer initiation. Instead our results are related and applicable to more general scenarios in population genetics, including situations in which a heterogeneous population is maintained through mutation-selection balance, or the case of Muller's ratchet when increasingly deleterious mutations become fixed \cite{muller1964relation}.

So far, most analytical investigations of stochastic tunneling \cite{komarova2003mutation,iwasa2004stochastic,nowak2004evolutionary,proulx2011rate} have been limited to considering transitions between homogeneous (or monomorphic) states of the population, as indicated in Fig.~\ref{fig:fig1}A. These investigations were performed assuming that cells proliferate according to the Moran process - a stochastic model of overlapping generations in which one cell division and one death event occur during each time step \cite{moran1962statistical}. \citeN{nowak2004evolutionary} analyzed the effect of the population size and mutation rates on the rate of appearance of a single cell with two mutations. These authors noted that for small, intermediate, and large populations, it takes two hits, one hit and zero hits, respectively, for a cell to accumulate two mutations. Here a hit is defined as a rate-limiting step, such as the appearance of an alteration when mutation rates are small. Considering fixation of cells with two mutations, \citeN{komarova2003mutation}, \citeN{iwasa2004stochastic} and \citeN{haeno2013stochastic} obtained explicit expressions for the tunneling rate. Subsequently, \citeN{iwasa2005population} used the assumption of independent lineages (i.e., individual lineages of cells harboring one mutation were considered to behave independently from each other) to compute the probability distribution for the time of emergence of a single second mutant in intermediate or large populations. \citeN{proulx2011rate} used a similar branching-process approach to derive a further tunneling rate.

Other types of dynamics such as the Wright-Fisher process have been studied as well, see e.g. \citeN{proulx2011rate}. In the Wright-Fisher process, cell generations are assumed to be non-overlapping, so that many birth and death events occur during each time step \cite{ewens2004population}. Using this process, \citeN{weinreich2005rapid} determined the critical population sizes for sequential fixation or stochastic tunneling, and \citeN{weissman2009rate} calculated the rate of tunneling as a function of the mutation rates, population size, and relative fitness of cells harboring one mutation. Finally, these results were extended to investigate the effects of recombination, or sexual reproduction, on the rate of stochastic tunneling \cite{weissman2010rate,lynch2010scaling,altland2011rare}. These authors found that the time to establishment of the double-mutant cells can be reduced by several orders of magnitude when sexual reproduction is considered \cite{weissman2010rate}. 

Recently, \citeN{proulx2012multiple} studied stochastic tunneling in a model which was not built upon the homogeneous-state approach. The author constructed a mutational network to study gene duplication. Although the setting of the model is very different from the setting we consider here, the underlying principles are similar.

The existing approaches for the Moran model in an asexual population provide accurate analytical approximations for a subset of the parameter space. We present a systematic overview of the scope of existing quantitative work. There are extensive regions of parameter space which, up to date, have been left unexplored by analytical approaches. These are predominantly situations in which the double mutant is not the most advantageous in the sequence. Before the double mutant reaches fixation, the population has to travel across a fitness hill or move constantly downhill in fitness space, as illustrated in Fig.~\ref{fig:fig1}B. The dynamics can then become trapped in quasi-equilibria -- a consequence of the mutation-selection balance \cite{crowkimura}. In these long-lived equilibria, the population is heterogeneous, and so previous approaches are not justified. Throughout our paper, these states are referred to as `metastable'. When these states exist, fixation is driven purely by demographic fluctuations. We address this regime based on ideas from mathematical statistical physics. Specifically we employ the Wentzel-Kramers-Brillouin (WKB) method to derive quantitative predictions for fixation times. Examples of using the WKB method to describe the escape from metastable states include the computation of mixing times in evolutionary games \cite{black2012mixing}, investigating extinction processes in coexisting bacteria \cite{lohmar2011switching} or predator-prey systems \cite{gottesman2012multiple}, and investigating epidemic models \cite{vanherwaarden1995stochastic,kamenev2008extinction,dykman2008disease,black2011wkb,billings2013intervention}.
In the presence of recombination, \citeN{altland2011rare} have shown that metastable states can appear when recombination rates are large, even if the double mutant is advantageous. The authors note that the time to escape across this `recombination barrier' increases exponentially with system size, as explained based on a WKB approach.

In this paper, we first classified the generic types of behavior that can occur in a population of cells which acquire two subsequent mutations in a Moran process: we determined when metastable states occur, when fixation is driven by intrinsic noise as opposed to deterministic flow, and where in parameter space fixation occurs in several subsequent hits. This classification was achieved by systematically studying the underlying deterministic dynamics of the process. We then obtained expressions for fixation times in parameter regimes which could not be captured by previous methods, i.e. regimes in which metastable states are found. We thus employed the WKB method to provide a more complete analytical description of the fixation dynamics in these parameter regimes. Our work fills the gap left by the existing literature and leads to a more comprehensive understanding of mutation acquisition and stochastic tunneling in evolving populations.

\section{The Model}
We considered a well-mixed, finite population of $N$ cells. Each cell can be of one of three possible types, labeled type 0 -- a wild-type cell harboring no mutations, type 1 -- a cell harboring one mutation, and type 2 -- a cell harboring two mutations. Initially, all cells are of type 0. The evolution of the population is determined by a Moran process \cite{moran1962statistical}. During each elementary time step of this process, a cell is randomly chosen to reproduce proportional to its fitness. In the same time step another cell is randomly removed, such that the total population size remains constant. The daughter cell can either inherit its type from the parent, or acquire a mutation during division. The relative fitness values of type-0, type-1 and type-2 cells are denoted by $r_0$, $r_1$ and $r_2$. Without loss of generality, we use $r_0=1$ throughout. The mutation rates $u_1$ and $u_2$ denote the probability that the daughter of a type-0 cell is of type-1, and the probability that the daughter of a type-1 cell is of type-2, respectively. We neglect all other combinations of mutations. The assumption of no back-mutation is commonly used in the population genetics literature \cite{ma2008infinite}. It is justifiable since the human genome is very large, $\sim3\times10^9$ base pairs, and the probability of mutating a specific base per cell division very small, $10^{-10}$ to $10^{-11}$ \cite{kunkel2000dna}. Therefore the chance of undoing a specific point mutation is vanishingly small. The probability that a second critical alteration occurs at a different locus is much higher.

In our model, finite populations will eventually reach a state in which all cells have acquired two mutations. This state is `absorbing', i.e. once this state has been reached, no further dynamics can occur. There are of course physical processes beyond the second mutation. In pre-cancerous tissues for example, there will be a finite probability that cells progress from this state to accumulate further changes (see Discussion). These processes are not the focus of our work though, and so are not included in the model.

Let us denote the number of type-0, type-1, and type-2 cells by $n_0$, $n_1$ and $n_2$, respectively; we have $N=n_1+n_2+n_3$. The transition rates for the Moran process are given by
\begin{eqnarray}
T^{1 \to 0} &=& \frac{(1-u_1)r_0\frac{n_0}{N}}{\rbar} n_1, \nonumber\\
T^{2 \to 0} &=& \frac{(1-u_1)r_0 \frac{n_0}{N}}{\rbar} n_2, \nonumber\\
T^{0 \to 1} &=& \frac{u_1 r_0\frac{n_0}{N}+(1-u_2)r_1\frac{n_1}{N}}{\rbar} n_0, \nonumber\\
T^{2 \to 1} &=& \frac{u_1 r_0\frac{n_0}{N}+(1-u_2)r_1\frac{n_1}{N}}{\rbar} n_2, \nonumber\\
T^{0 \to 2} &=& \frac{u_2 r_1\frac{n_1}{N}+r_2\frac{n_2}{N}}{\rbar} n_0, \nonumber\\
T^{1 \to 2} &=& \frac{u_2 r_1\frac{n_1}{N}+r_2\frac{n_2}{N}}{\rbar} n_1.
\label{eq:transitionrates}
\end{eqnarray}
The quantity $\rbar=(r_0n_0+r_1n_1+r_2n_2)/N$ is the average fitness in the population. The transition labeled `$i \to j$' represents a process in which a cell of type $i$ is replaced by a cell of type $j$. In a process labeled `$1\to 0$', for example, the state of the population changes from ${(n_0,n_1,n_2)}$ to state ${(n_0+1,n_1-1,n_2)}$.
As an example, the first reaction rate, $T^{1 \to 0}$, in Eq.~(\ref{eq:transitionrates}) can be broken down as follows: a type-0 cell is chosen to reproduce at rate $(n_0/N)\times (r_0/\rbar)$. The offspring does not mutate with probability $(1-u_1)$. Finally, a type-1 cell is chosen to be removed at rate $n_1$. The rates for the other processes can be interpreted analogously. We choose a continuous-time setup, and correspondingly all rates in Eq. (\ref{eq:transitionrates}) scale linearly in the population size $N$. Simulations are carried out using a standard Gillespie algorithm \cite{gillespie1977exact}, and times are measured in cellular generations.

This process is described exactly by a master equation \cite{vankampen2007stochastic}, which governs the behavior of the probability, $P_{\vec{n}}(t)$, that the population is in state $\vec{n}=(n_0,n_1,n_2)$ at time $t$, and is given by 
\begin{equation}
\dot{P}_{\vec{n}}(t) = \sum_{\vec{v}} \left[T^{\vec{v}}_{\vec{n}-\vec{v}} P_{\vec{n}-\vec{v}}(t) - T^{\vec{v}}_{\vec{n}} P_{\vec{n}}(t)\right].
\label{eq:masterequation}
\end{equation}
The vector $\vec{v}$ indicates a change in the composition of the population due to the corresponding reaction, and $\dot{P}_{\vec{n}}(t)$ represents the partial derivative of $P_{\vec{n}}(t)$ with respect to time. This equation states that the probability that the population is in state $\vec{n}$ at time $t$ increases due to transitions into the state $\vec{n}$ and decreases due to transitions out of the state $\vec{n}$. 

The master equation contains full information about the stochastic population dynamics. In particular, the detailed statistics of the population at any time can be derived from it, and it captures effects driven by intrinsic noise, such as extinction and fixation. Obtaining a full solution of the master equation is difficult or impossible though in all but the simplest of cases. As a starting point, it is often useful to first consider the deterministic limit of infinite populations. In this limit, the distribution $P_{\vec{n}}(t)$ is sharply peaked around its average, and so the dynamics reduces to a set of equations for this mean. This approach does not capture any of the stochastic effects. However, the types of stochastic trajectories that can be observed for different parameter sets are, to some extent, set by the underlying deterministic dynamics. We thus first analyze the deterministic limit of the model.

\section{Deterministic analysis and types of stochastic behavior}\label{sec:deterministic}
In the limit $N \to \infty$, the population evolves according to a deterministic set of equations. Writing $x_i= \lim_{N\to \infty}n_i/N$, we have the relation ${x_0+x_1+x_2=1}$, and the average fitness is given by ${\rbar=r_0 x_0 + r_1 x_1 + r_2 x_2}$.  The equations governing the dynamics of the population are then
\begin{eqnarray}
\rbar \dot{x}_0 &=& \bigl[(1-u_1)r_0-\rbar\bigr]x_0, \nonumber\\
\rbar \dot{x}_1 &=& u_1 r_0 x_0 +\bigl[(1-u_2)r_1 -\rbar\bigr]x_1, \nonumber\\
\rbar \dot{x}_2 &=& u_2 r_1 x_1 +(r_2-\rbar)x_2.
\label{eq:determinsticequations}
\end{eqnarray}
These equations can be derived systematically using a system-size expansion of the master equation~(\ref{eq:masterequation}), see e.g. \citeN{vankampen2007stochastic}. Note that $x_i$ refers to the relative concentration of cells of type $i$ in the population (not to be confused with the probability of being found in a homogeneous state of type-$i$ cells as studied in \citeN{iwasa2004stochastic} and \citeN{haeno2013stochastic}). For example, $x_0=x_1=x_2=1/3$ would indicate that the population is in a state in which all three types are present with equal numbers. 

\begin{figure}
\begin{center}
\includegraphics*[width=\textwidth]{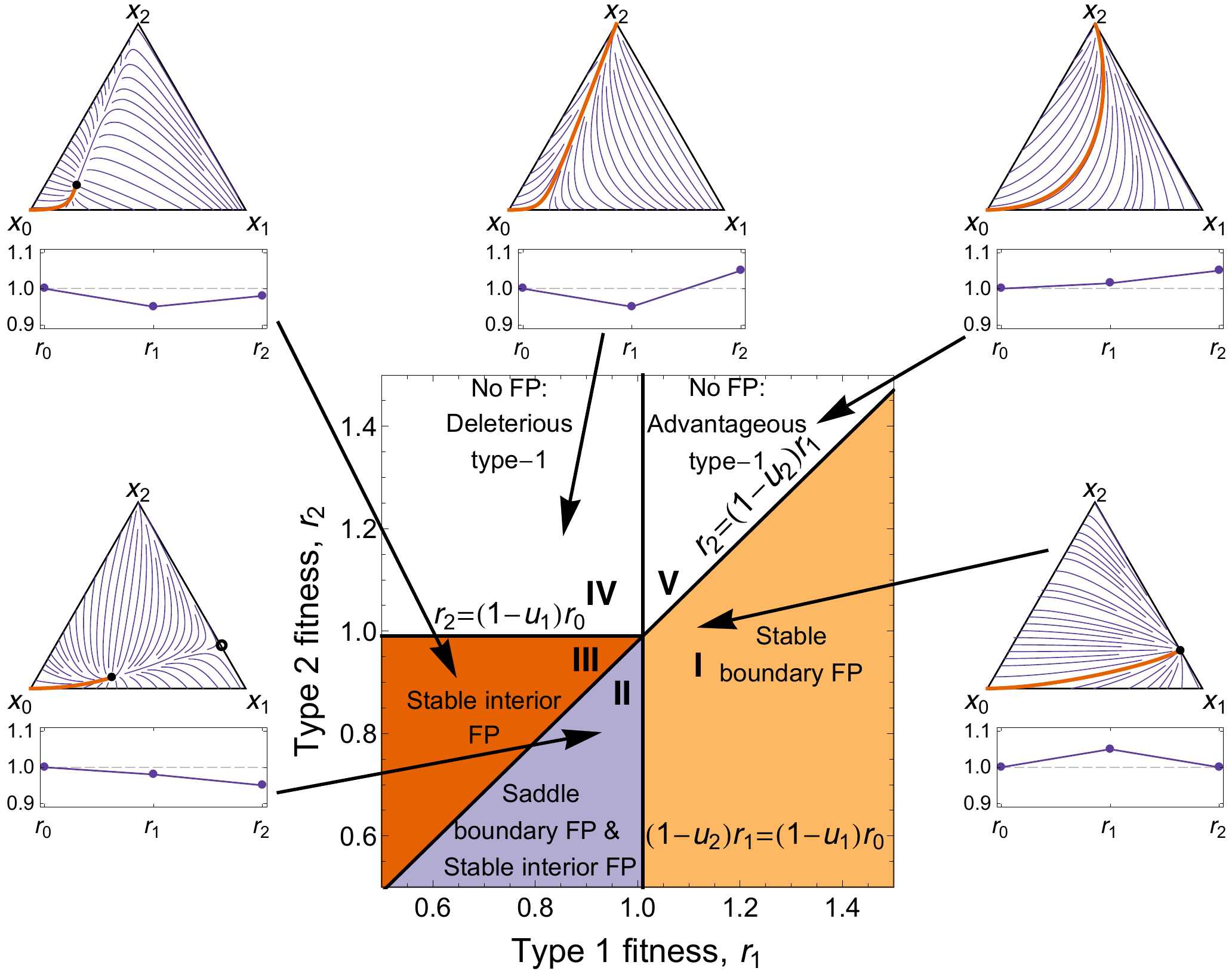}
\end{center}
\caption{\textbf{Phase diagram with satellite diagrams showing the deterministic flow.}
Fixed points are found on the 1--2 boundary and/or in the interior of the simplex. Boundary fixed points occur when $r_2<(1-u_2)r_1$ (Regions I and II, mutation-selection balance between types $1$ and $2$). Stable interior fixed points occur when $r_2<(1-u_1)r_0$ and $(1-u_2)r_1<(1-u_1)r_0$ (Regions II and III, mutation-selection balance between all three types). No fixed points are found in Regions IV and V (beneficial type-$2$ mutation). The satellite diagrams show the deterministic flow, Eq.~(\ref{eq:determinsticequations}). Thick (red) lines show the deterministic flow from the all-wild-type initial condition. Solid circles indicate stable fixed points, and the empty circle for Region II corresponds to a saddle point which is stable along the 1--2 boundary. In all cases, the point $x_2=1$ is an absorbing state and is therefore a fixed point as well. Below each simplex we illustrate the shape of the fitness landscape which generates each type of behavior.}
\label{fig:fig2}
\end{figure}

Given the relation $x_0+x_1+x_2=1$, the dynamics only has two independent degrees of freedom. Time courses of the system can hence be thought of as a trajectory in a `concentration simplex', as depicted in the satellite diagrams of Fig.~\ref{fig:fig2}. Each point in the simplex represents one particular state $(1-x_1-x_2, x_1,x_2)$ of the population. At points in the interior of the simplex all three types of cells are present in the population ($x_i>0$ for $i=0,1,2$). Points on the edges of the simplex represent states in which one of the three types is not present, for example $x_0=0$ for points along the edge connecting the lower-right corner of the simplex with the upper corner. We will refer to this as the 1--2 edge in the following, and similarly for the other edges. The three corners of the simplex represent the homogeneous states, i.e. $x_0=1$ (lower left corner), $x_1=1$ (lower right) and $x_2=1$ (upper corner).

The deterministic equations~(\ref{eq:determinsticequations}) have a trivial fixed point (a point $(1-x_1^*-x_2^*,x_1^*,x_2^*)$ at which $\dot{x}_i=0$ for all $i=0,1,2$) at $x_2=1$, corresponding to the absorbing state. The equations can have a further zero, one, or two non-trivial fixed points, depending on the values of the fitness parameters and the mutation rates. These fixed points correspond to points at which mutation and selection balance \cite{crowkimura}. Each fixed point can either be stable (i.e. attracting from all directions) or a saddle (attracting from some directions, repelling in others). The system is found not to contain any fully repelling fixed points (Appendix~\ref{app:FP}). Fig.~\ref{fig:fig2} shows the deterministic dynamics in different parameter regimes, indicated as Regions I to V. Below we discuss the stochastic behavior in each of these parameter regimes.

{\em Region I (mutation-selection balance between type-$1$ and type-$2$ cells):}\\
In Region I, the deterministic dynamics flows towards a fixed point on the 1--2 edge of the concentration simplex ($x_0=0$).
The type-0 cells have the lowest fitness, and are deterministically lost by selection.
The fixed point is a consequence of mutation-selection balance between type-1 and type-2 cells \cite{crowkimura}. Writing $r_2=(1-s)r_1$, the existence condition for the equilibrium ($r_2<(1-u_2)r_1$) reduces to $u_2<s$. It is a well-known result from population genetics that this condition prevents the deterministic loss of the type-1 cells \cite{crowkimura}. The deterministic system gets stuck at this fixed point, but a finite population will eventually reach fixation in the all-2 state. 
At large but finite population sizes, the stochastic dynamics are expected to approximately follow the deterministic path shown in Fig.~\ref{fig:fig2} such that type-0 cells quickly become extinct. The lack of backwards mutations means the population cannot depart from this edge and the problem reduces to one degree of freedom. The mutation-selection balance maintains the heterogeneous population state of type-1 and type-2 cells. The intrinsic noise then has to drive the system from this metastable state into the absorbing all-2 state against the direction of selection. Fixation times are expected to grow exponentially with the population size \cite{antal2006fixation,mobilia2010oscillatory,altland2011rare}.

\begin{figure}
\begin{center}
\includegraphics*[width=0.8\textwidth]{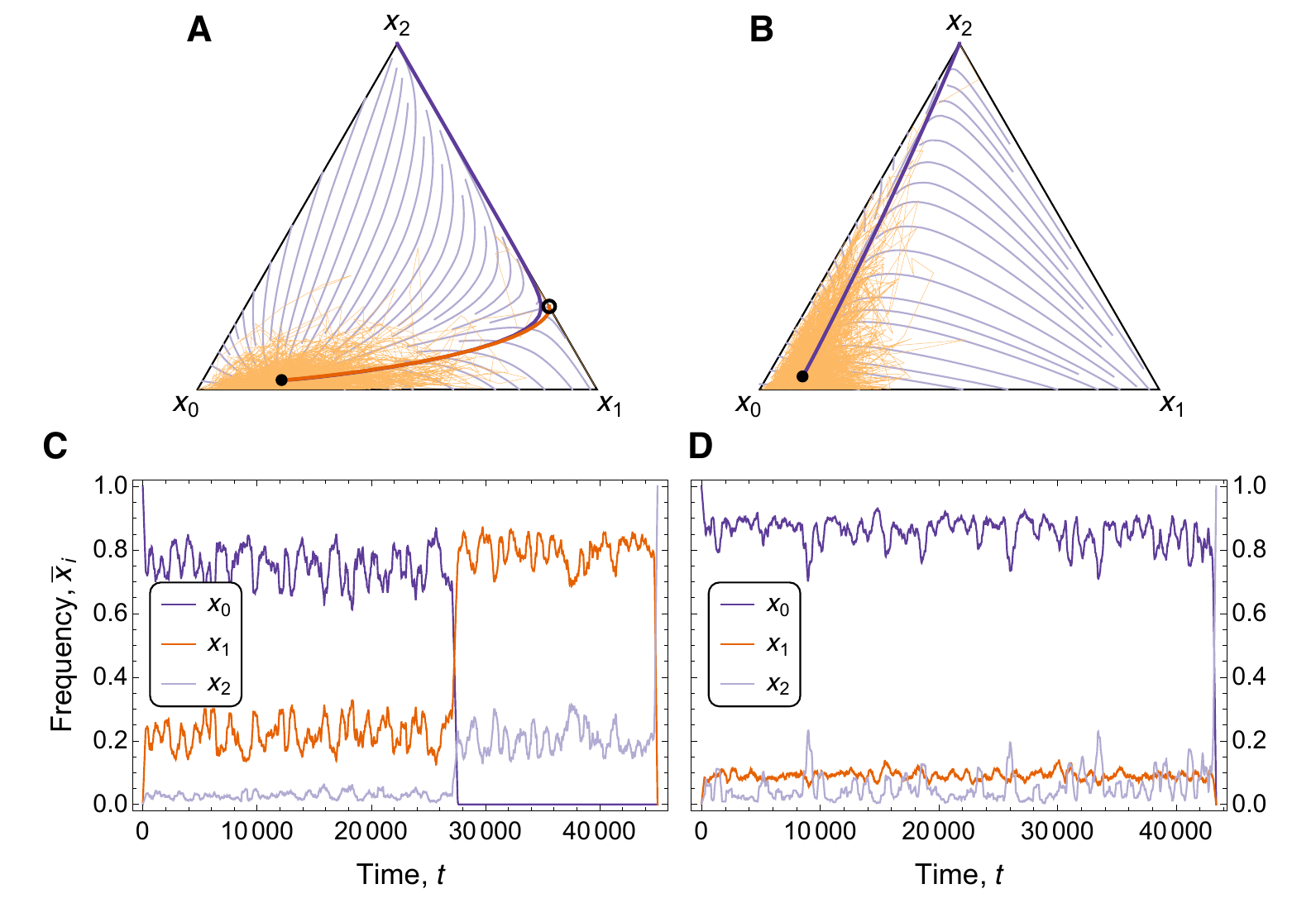}
\end{center}
\caption{\textbf{Routes to fixation in cases with a stable interior fixed point.} 
\textbf{A} The concentration simplex for a combination of parameters in Region II ($r_1=0.96$, $r_2 = 0.91$, $u_1 = 0.01$, $u_2 = 0.01$). The red line is the dominant path for sequential extinction, as obtained from the WKB analysis (see text). The thick purple line indicates the dominant path for simultaneous extinction of types 0 and 1, which is rare. Faint lines indicate the deterministic flow [Eq.~(\ref{eq:determinsticequations})]. The thin orange line represents the trajectory of a single stochastic simulation ($N=300$).
\textbf{B} Dominant trajectory, flow lines, and stochastic trajectory for a combination of parameters in Region III ($r_1 = 0.90$, $r_2 = 0.97$, $u_1 = 0.01$,  $u_2 = 0.01$). Cells of types 0 and 1 go extinct (essentially) at the same time $t\approx 44,000$. The dominant trajectory as obtained from the WKB calculation (thick purple line) runs directly into the all-2 corner of the concentration simplex.
\textbf{C} The concentrations of the three types of cells as a function of time. These are obtained from the same stochastic simulation as shown in the simplex \textbf{A}. A moving average has been taken to improve clarity. As seen from the data, cells go extinct sequentially: initially all three types are present, cells of type 0 go extinct at time $t\approx 28,000$, cells of type 1 go extinct at time $t\approx 45,000$.
\textbf{D} The concentrations (moving average) of the three types of cells as a function of time for the stochastic simulation shown in the simplex \textbf{B}, showing the simultaneous extinction of types 0 and 1. } 
\label{fig:fig3}
\end{figure}

{\em Region II (mutation-selection balance between all three types, and, separately, between types $1$ and $2$):}\\
In Region II, the deterministic flow from the all-wild-type state is towards a stable fixed point in the interior of the simplex. This point corresponds to the mutation-selection balance point of all three species. There is a second fixed point located on the 1--2 edge, which corresponds to mutation-selection balance between types 1 and 2 in the absence of type-0 cells (analogous to Region I). As type-0 cells have the highest fitness in this regime, selection is directed away from the 1--2 edge. Thus the fixed point on this edge is a saddle.
As before the stochastic dynamics in finite populations will reach the all-2 state eventually. The population will closely follow the deterministic trajectory (see Fig.~\ref{fig:fig2}) before reaching the metastable state about the stable interior fixed point.
Here the mutation-selection balance maintains the heterogeneous state with all three species present.
The population will fluctuate about this fixed point until it eventually overcomes the adverse selection and escapes. There are two possibilities for the subsequent behavior:
(i) Type-0 cells become extinct and the population settles into the metastable state on the 1--2 edge. Intrinsic fluctuations enable the population to overcome the adverse selection along the edge and reach the absorbing all-2 state. This corresponds to sequential extinction, first of type-0 cells, then of type-1 cells. This process is equivalent to a minimal model of Muller's ratchet \cite{muller1964relation}, in which the most advantageous phenotypes are sequentially lost. A trajectory of this type is illustrated in Figs.~\ref{fig:fig3}A and \ref{fig:fig3}C;
(ii) Cells of type 0 and type 1 can in principle go extinct (almost) simultaneously. The trajectory of the system then hits the 1--2 edge infinitesimally close to the all-2 corner of the simplex (here `infinitesimally close' means a distance of order $1/N$ away from the upper corner). It does not become trapped in the metastable state located on the 1--2 edge. In numerical simulations (data not shown) we find that this second path is realized only very rarely, and so our mathematical analysis of Region II below focuses on sequential extinction.

{\em Region III (mutation-selection balance between all three types):}\\
In Region III the deterministic dynamics has a single stable fixed point in the interior of the concentration simplex. This point again corresponds to the mutation-selection balance point of all three species.
Large, but finite populations will behave as discussed in case (ii) for Region II. They will initially become trapped in the metastable state about the mutation-selection balance point, before intrinsic fluctuations eventually drive the system to the absorbing all-2 state.
In Region III, type-0 cells and type-1 cells go extinct at essentially the same time. If the type-0 cells become extinct first, then type-1 cells quickly become extinct as selection along the 1--2 edge is directed towards the absorbing state ($r_2>(1-u_2)r_1$). This is illustrated in Figs.~\ref{fig:fig3}B and \ref{fig:fig3}D.

{\em Regions IV and V (beneficial type-$2$ mutation):}\\
In a subset of the parameter space, shown as Regions IV and V in Fig.~\ref{fig:fig2}, the deterministic flow from the all-wild-type state is directly to the absorbing all-2 state. For such model parameters we expect that fixation in finite populations will be quick as type-2 cells are favored by selection (and mutation).
These scenarios agree with the theory of natural selection, in which the populations fitness increases over time \cite{fisher1930genetical}. In Region IV this is achieved by crossing a fitness valley, and in Region V it is achieved by sequentially selecting the most advantageous phenotype.

\bigskip

\begin{figure}
\begin{center}
\includegraphics*[width=0.8\textwidth]{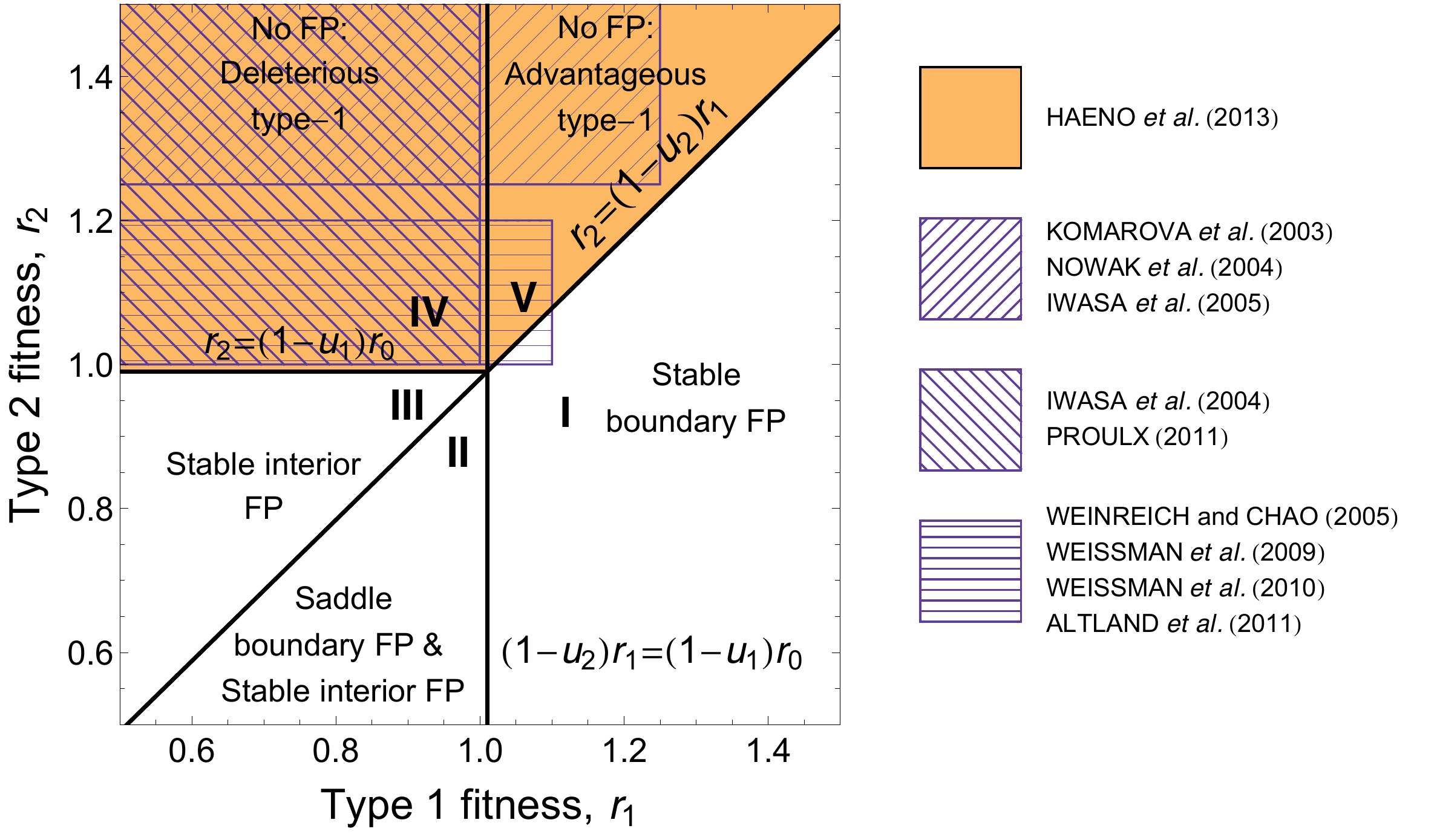}
\end{center}
\caption{\textbf{Phase diagram highlighting previous work.}
The colored area roughly corresponds to the regions in \protect\citeN{haeno2013stochastic} in which the fixation probability is successfully predicted (See Fig. 6 in \protect\citeN{haeno2013stochastic}). 
The southwest-northeast striped region, with $r_1$ deleterious or slightly advantageous, and $r_2$ very advantageous, is the approximate region of interest of the studies \protect\citeN{komarova2003mutation}, \protect\citeN{nowak2004evolutionary}, and \protect\citeN{iwasa2005population}. These studies focused on the time to emergence of a single type-2 cell.
The northwest-southeast striped region, with $r_1$ neutral or deleterious, and $r_2$ advantageous, is approximately the region of interest of \protect\citeN{iwasa2004stochastic} and \protect\citeN{proulx2011rate}. These studies were concerned with computing fixation times of the advantageous type-2 cells and rely on the assumption that the number of type-1 cells is small.
Finally, the horizontal striped region approximately corresponds to the literature of crossing fitness valleys, notably \protect\citeN{weinreich2005rapid}, \protect\citeN{weissman2009rate}, \protect\citeN{weissman2010rate}, \protect\citeN{lynch2010scaling}, and \protect\citeN{altland2011rare}. These studies are concerned with $1/N < r_2-r_0 \ll 1$.}
\label{fig:fig4}
\end{figure}

Fig.~\ref{fig:fig4} illustrates which parameter regimes have previously been studied in the stochastic tunneling literature. These existing studies almost exclusively focus on Regions IV and V, i.e. cases in which fixation is driven not primarily by demographic noise, but by the underlying deterministic flow. As mentioned above fixation is typically fast in Regions IV and V. Based on similar studies in evolutionary game theory one would expect the fixation time to grow logarithmically with the population size, $\tau\sim\ln N$ \cite{antal2006fixation}, and this is indeed what we find in simulations (data not shown). 
The regions containing non-trivial fixed points are largely unexplored by previous investigations. Fixation is controlled by stochastic effects so that fixation times are large and broadly distributed. As we will discuss below, fixation times grow exponentially with the population size in such cases. This is perfectly in-line with the findings of \citeN{haeno2013stochastic}, who point out that fixation in these regions takes a very long time. Efficient measurements of fixation time in simulations are hence difficult. Methods which require the numerical solutions of, for example, the backward Fokker-Planck equation or a backward master equation reach their limits here as well \cite{haeno2013stochastic}. The contribution of our work is to analyze precisely these previously inaccessible cases. We compute the fixation properties of systems in which the underlying deterministic flow has one or more attracting fixed points away from the absorbing all-2 state.

\section{WKB analysis}
Let us now analyze the dynamics in Regions I, II and III, i.e. situations in which the deterministic dynamics has one or two non-trivial fixed points. 
In large but finite populations these fixed points correspond to metastable states in which the effect of mutation and selection balance. The aim of the following analysis is to calculate the rate at which the population will escape this state and enter the absorbing state in which all cells harbor two mutations. To proceed with the analysis we make the following simplifying assumptions, which are justified by the previous deterministic analysis:
\begin{enumerate}
\item We assume that the population first settles into a distribution about the mutation-selection balance point. This distribution is calculated below.
\item We assume that the population will `leak' into the absorbing state on a very long timescale from this distribution. With this assumption we can also say that the time taken for the population to reach the metastable state is negligible when compared to the escape time.
\end{enumerate}
With these assumptions we can compute (from the master equation (\ref{eq:masterequation})) the distribution about the mutation-selection balance point and the escape rate. These assumptions (and hence the subsequent analysis) require the selective pressure to be greater than the effect of noise, such that the metastable states are long-lived. For this reason, the approach described here is only valid for large $N$ which satisfy this condition (the minimum value of $N$ for which our analysis is valid is dependent on the remaining model parameters, but comparisons with simulation results show it is accurate for $N\ge100$).

Mathematically we formulate the problem as follows: the shape of the distribution about the mutation-selection balance point is given by $\pi_{\vec{n}}$, and is henceforth referred to as the quasi-stationary distribution (QSD) in line with existing literature \cite{assaf2010extinction}. The mean time taken to escape from the metastable state, $\tau$, is much greater than the time taken to initially reach the metastable state $t_r$, i.e. ${\tau \gg t_r}$. Provided this condition holds, we can assume that after a short time the probability to find the population in state $\vec{n}$ is given by
\begin{equation}
P_{\vec{n} \ne \vec{n}_{\rm abs}}(t \gg t_r) \simeq \pi_{\vec{n}}\, e^{-t/\tau},~~P_{\vec{n}_{\rm abs}}(t \gg t_r) \simeq 1-e^{-t/\tau}.
\label{eq:quasistationaryapprox}
\end{equation}
The exponential decay factor, $e^{-t/\tau}$, describes the `leaking' process from the metastable state into the absorbing state, $\vec{n}_{\rm abs}=(0,0,N)$. The second equation follows from normalization.

To find the mean fixation time of the type-2 cells, we substitute Eq.~(\ref{eq:quasistationaryapprox}) into the master equation~(\ref{eq:masterequation}) to obtain the quasi-stationary master equation (QSME)
\begin{equation}
0\approx -\frac{1}{\tau}\pi_{\vec{n} \ne \vec{n}_{\rm abs}} =
\sum_{\vec{v}} T^{\vec{v}}_{\vec{n}-\vec{v}} \pi_{\vec{n}-\vec{v}} - T^{\vec{v}}_{\vec{n}} \pi_{\vec{n}}. \label{eq:QSME}
\end{equation}
For $\vec{n}=\vec{n}_{\rm abs}$ (the absorbing state) we have
\begin{equation}
\frac{1}{\tau}=\sum_{\vec{v}} T^{\vec{v}}_{\vec{n}_{\rm abs}-\vec{v}} \pi_{\vec{n}_{\rm abs}-\vec{v}}
=T^{1\to2}_{(0,1,N-1)}\pi_{(0,1,N-1)} + T^{0\to2}_{(1,0,N-1)}\pi_{(1,0,N-1)}, \label{eq:timescaleansatz}
\end{equation}
where we have used $T^{\vec{v}}_{\vec{n}_{\rm abs}}=0$.
Hence if we find the QSD $\pi_{\vec{n}}$ by solving the QSME (\ref{eq:QSME}), we can determine the mean fixation time, $\tau$, and the probability to have reached fixation by time $t$, $P_{\vec{n}_{\rm abs}}(t)=1-e^{-t/\tau}$.

By separating variables in Eq.~(\ref{eq:quasistationaryapprox}), we have reduced the complexity of the master equation~(\ref{eq:masterequation}) (time does not feature in the QSME~(\ref{eq:QSME})). If we now replace the discrete variables $\vec{n}$ with continuous variables $\vec{x}=\vec{n}/N$, we further reduce the complexity. This continuous approximation is valid as we have already stated that we require $N$ to be large.
We now employ the WKB ansatz to represent the QSD as
\begin{equation}
\pi_{N\vec{x}} = \psi(\vec{x}) = C\exp\left[-N S(\vec{x}) - S_1(\vec{x})+\mathcal{O}(N^{-1})\right], \label{eq:WKBansatz}
\end{equation}
where $S(\vec{x})\sim\mathcal{O}(1)$ is known as the \emph{action}, and $\exp[-S_1(\vec{x})]$ ($S_1\sim\mathcal{O}(1)$) is the so-called \emph{amplitude} \cite{assaf2010extinction}. We have introduced $C$ as a normalization constant. To find the QSD, we follow e.g. \citeN{assaf2010extinction} and expand the QSME~(\ref{eq:QSME}) in powers of $N^{-1}$. Further analysis can be carried out if the QSME has only one variable. This is relevant in Regions I and II of our model, where the population must escape from a metastable state on the 1--2 edge of the concentration simplex. An example of the QSD obtained from this procedure is shown in Fig.~\ref{fig:fig5}.

Escape from an interior metastable state can also be studied using the WKB approach. The QSME then retains two degrees of freedom, and explicit expressions for the action and escape time cannot be obtained. In the following we briefly describe the main steps for each of the two situations.

\subsection{Analysis of mutation-selection balance between types $1$ and $2$ \label{ssec:analytic}}
We first consider the case in which the population must escape from a metastable state on the 1--2 edge (Regions I and II). As the population cannot depart from the 1--2 boundary, the system then reduces to one degree of freedom. We parameterize the system in terms of $n_1$, such that $n_1=0$ is the absorbing state (all cells of type 2) and $n_2=N-n_1$. The analysis now closely follows the work of \citeN{assaf2010extinction}, specifically their scenario A. The outcome of the analysis are expressions for the action $S(x_1)$, $S_1(x_1)$ (determined up to an additive constant), and the normalization constant $C$. With this we find an explicit expression for the mean escape time, $\tau$, from the metastable state. In Region I in which there exists just a single deterministic fixed point, this time corresponds to the mean fixation time, as documented above. The fixation time in Region II is discussed in the next section. This analysis is covered in detail in Appendix~\ref{app:sol}, we here only list the main steps and the result of the solution procedure:

\begin{enumerate}
\item Substitute the WKB ansatz~(\ref{eq:WKBansatz}) into the QSME~(\ref{eq:QSME}), and then expand in powers of $N^{-1}$. The equation at leading order in $N$ can be solved to find the action $S(x_1)$. The equation at next-leading order can be solved to find the function $S_1(x_1)$. The QSD~(\ref{eq:WKBansatz}) is normalized by considering a Gaussian approximation of $\psi(x_1)$. Note that with expressions for $S$, $S_1$ and $C$, we have derived the distribution about the mutation-selection balance point. An example of this is plotted in Fig.~\ref{fig:fig5}.

\item Find a so-called boundary-layer solution by recursively solving the QSME~(\ref{eq:QSME}) for ${0<n_1 \lesssim N^{1/2}}$, using Eq.~(\ref{eq:timescaleansatz}) to express the solution as a function of the expected escape time $\tau$.

\item Matching the recursive boundary-layer solution and the WKB solution in the domain $0<n_1<N^{1/2}$ gives the expected escape time from the metastable state to the absorbing state,
\begin{eqnarray}
\tau &=& \sqrt{\frac{2\pi r_1 r_2}{N u_2}} \frac{(1-u_2)(r_1-r_2)}{[(1-u_2)r_1-r_2]^2} \nonumber\\
&&\times \exp\left\{
N\left[\frac{u_2 r_1}{r_2-u_2 r_1} \ln\left(\frac{u_2(1-u_2)r_1^2}{r_1-r_2}\right)
+\ln\bigl((1-u_2)r_1\bigr)
-\frac{r_2}{r_2-u_2 r_1}\ln r_2 
\right]\right\}.~~
\label{eq:fixationtime}
\end{eqnarray}
\end{enumerate}

\subsection{Analysis of mutation-selection balance between all three types}
When a fixed point exists away from the state-space boundaries, as found in Regions II and III in Fig.~\ref{fig:fig2}, the solution procedure described above is no longer viable; the equations recovered from step (i) above can no longer be integrated directly subject to the appropriate boundary conditions (see Appendix~\ref{app:num}). Instead, minimum-action Hamiltonian trajectories have to be computed. Problems of this type are usually tackled in one of two ways: firstly, the equations of motion could be integrated using a shooting method to find the optimal (most likely) trajectory with a given final point \cite{kamenev2008extinction,black2011wkb,gottesman2012multiple}; secondly, the equations can be integrated using an iterative scheme which converges to the optimal trajectory \cite{lohmar2011switching} connecting given start and end points. We found that the second method quickly converges for our problem, so results presented in the following use this method. The details of the procedure are described in Appendix \ref{app:num}.

Using this method we calculated the action, $S$, accumulated over the trajectory from the stable interior fixed point to the boundary saddle fixed point in Region II, or directly to the absorbing state in Region III. We then used a known result presented by \citeN{vanherwaarden1995stochastic} for the escape time from a metastable state,
\begin{equation}
\tau \sim \frac{C}{\sqrt{N}} e^{N S}, \label{eq:taufit}
\end{equation}
where the constant $C$ can be found by fitting to simulation data. This expression has the same functional dependence on $N$ as the one given in Eq.~(\ref{eq:fixationtime}).

In principle, one could numerically find the action of a trajectory starting at the stable fixed point and finishing at a general position $\vec{x}$ in concentration space. This would yield the values $S(\vec{x})$, and the distribution about the metastable state is then given by $\psi(\vec{x})\sim\exp[-NS(\vec{x})]$. This very technical analysis is tedious and beyond the scope of the present paper.

\section{Results}

We analyze the results separately for each region (I, II and III) of parameter space. In particular we discuss the implications the model parameters have for the probability with which tunneling occurs, and for the fixation time of type-2 cells.

\subsection{Region I}
In this region, type-1 cells have a fitness advantage over both type-0 and type-2 cells, such that the fitness landscape has an intermediate maximum (a fitness `hill'; see Fig.~\ref{fig:fig2}). As a result, type-0 cells are deterministically lost and the population relaxes to the QSD (i.e. the mutation-selection equilibrium) on the 1--2 edge, as described above, and probability slowly leaks into the absorbing state.

\begin{figure}
\begin{center}
\includegraphics*[width=0.7\textwidth]{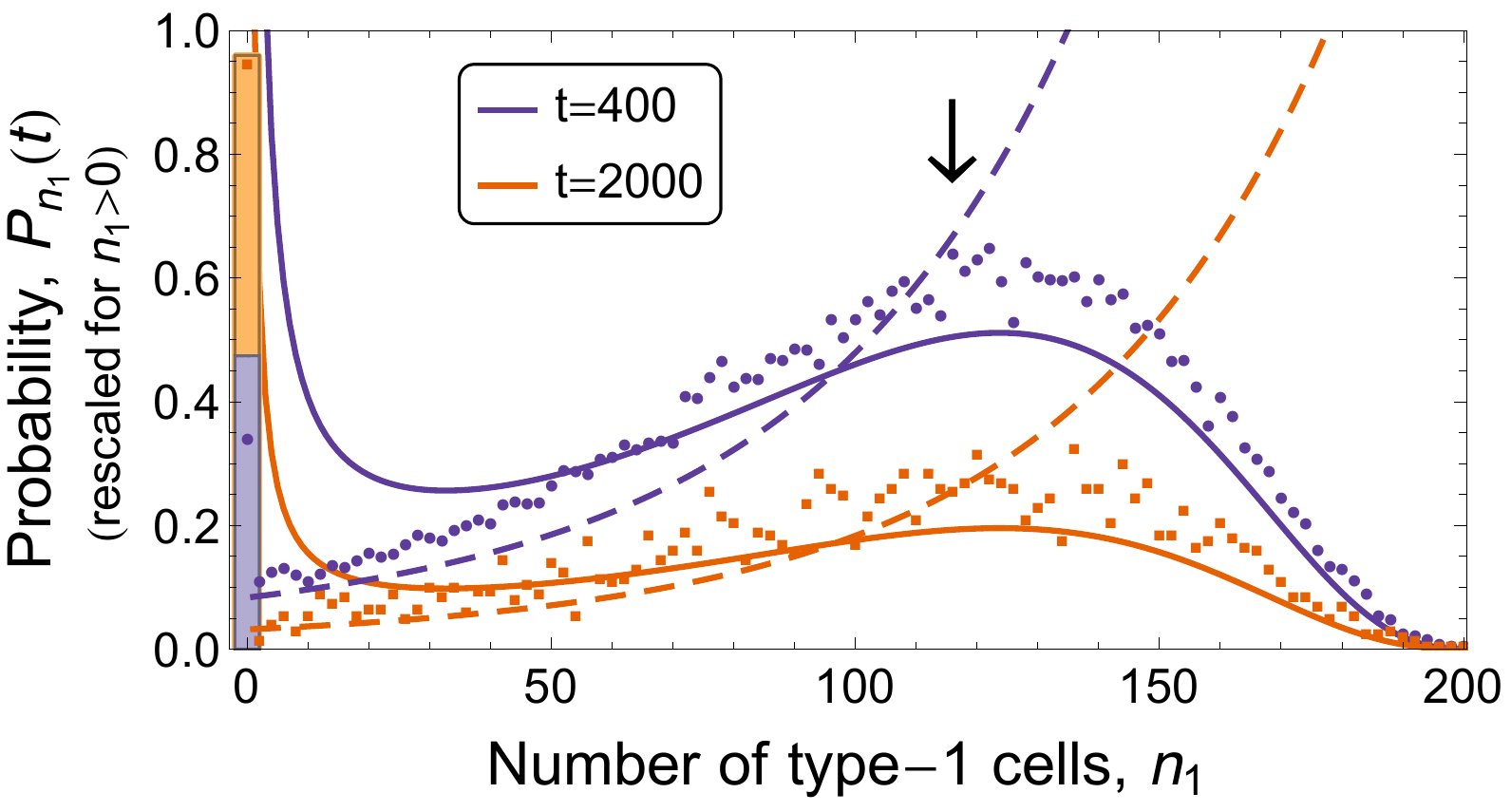}
\end{center}
\caption{\textbf{Quasi-stationary distribution compared with simulation results.}
We show the approximation Eq.~(\ref{eq:quasistationaryapprox}) (solid lines; filled bars for $n_1=0$) against the distribution of states $n_1$ obtained from an ensemble of $10^5$ simulation runs (symbols). Here the metastable state is located along the 1--2 boundary (Region I: here the solution is obtained from the theory with $S(x_1)$ and $S_1(x_1)$ given by Eq.~(\ref{eq:SS1}), and the normalization $C$ given by Eq.~(\ref{eq:norm})). Dashed lines correspond to the boundary-layer solution of the master equation, valid for $0 \lesssim n_1 < N^{1/2}$ [Eq.~(\ref{eq:recursive})]. Note that the distributions away from $n_1=0$ have been re-scaled by factors of $100$ for $t=400$ and $500$ for $t=2000$ for optical convenience. The arrow indicates the location of the fixed point. Parameters are $r_0=1.0$, $r_1=1.05$, $r_2=1.0$, $u_1=1\times10^{-2}$, $u_2=2\times10^{-2}$, $N=200$.}
\label{fig:fig5}
\end{figure}

To test the accuracy of this approach, in Fig.~\ref{fig:fig5} we compare the QSD measured in simulations with the theoretical approximation. The data in the figure reveals good agreement between theory and model simulations for $N^{1/2} \lesssim n_1 \le N$, and for $n_1=0$. In the region just above $n_1=0$ the agreement between the theoretical result for $P_{n_1}$ and simulation data breaks down; here the theoretical value from Eq.~(\ref{eq:WKBansatz}) diverges. This is a result of taking the limit $1/\tau \to 0$ in the QSME~(\ref{eq:QSME}), in particular for the case $n_1=0$ which corresponds to $1/\tau = T_{(0,1,N-1)}^{1 \to 2} \pi_1$. The value of $\tau$ is crucial to determining the value of $\pi_1$.  When calculating the mean fixation time, $\tau$, we circumvented this known problem by considering a so-called `boundary-layer' approach \cite{assaf2010extinction}. The boundary-layer solutions (dashed lines in Fig.~\ref{fig:fig5}; for details of the calculation see Appendix~\ref{app:sol}) show better agreement with simulation results close to $n_1=0$ than the QSD obtained from the WKB ansatz (solid lines).

\begin{figure}
\begin{center}
\includegraphics*[width=\textwidth]{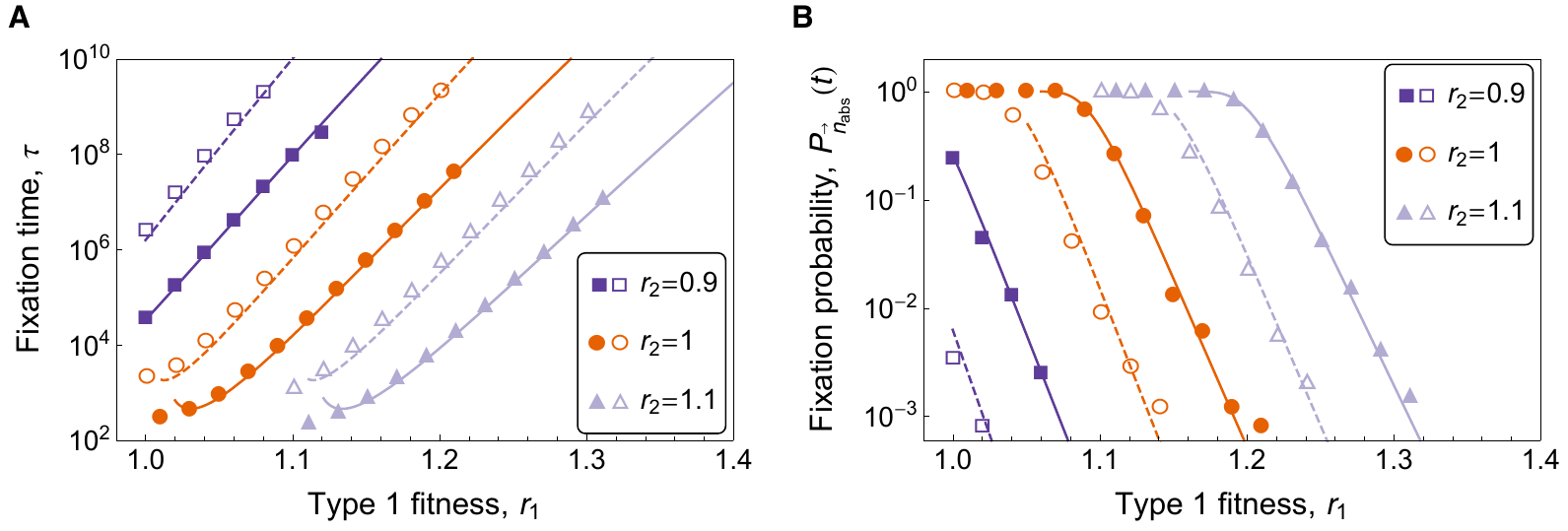}
\end{center}
\caption{\textbf{Results for Region I.}
\textbf{A} Mean fixation time of type-2 cells in Region I. Symbols correspond to mean fixation time from simulations of the model (averaged over an ensemble of $2$--$5\times10^3$ samples) initiated in the all-wild-type state. Shape of symbol indicates fitness of type-2 cells (see legend); filled symbols are for $u_1=u_2=10^{-2}$, empty symbols are for $u_1=u_2=10^{-3}$. Solid lines (high mutation) and dashed lines (low mutation) are the WKB prediction for fixation time, Eq.~(\ref{eq:fixationtime}). The approximation breaks down when ${(1-u_2)r_1\simeq r_2}$, which is when the fixed point approaches the absorbing state. 
\textbf{B} Fixation probability of type-2 cells in Region I, evaluated at time $t=10^4$. Lines correspond to the WKB prediction Eqs.~(\ref{eq:quasistationaryapprox}) and (\ref{eq:fixationtime}), and colors and symbols follow the same convention as in panel \textbf{A}.
Remaining parameters are $r_0=1.0$ and $N=100$.}
\label{fig:fig6}
\end{figure}

Results for the mean fixation time in Region~I are shown in Fig.~\ref{fig:fig6}A. In Fig.~\ref{fig:fig6}B, we plot the probability that type-2 cells have reached fixation by time $t=10^4$ (including fixation earlier than that). We refer to this quantity as the fixation probability.
Fixation times are shown to increase exponentially with $r_1$. This is a consequence of the increasing height of the selection `barrier' which must be overcome for type-2 cells to reach fixation. Also, increasing $r_1$ pushes the mutation-selection balance point towards the all-1 state. This results in a further increase in fixation time (or decrease in fixation probability). As the metastable state approaches the all-1 state, the probability of the population reaching the all-1 state due to demographic fluctuations increases. Thus increasing $r_1$ decreases the probability of tunneling.

Increasing the fitness of type-2 cells, on the other hand, pushes the metastable state closer to the absorbing state. This leads to a significant reduction in the fixation time (increase in fixation probability) as also shown in Fig.~\ref{fig:fig6}A (\ref{fig:fig6}B). Increasing the mutation rate $u_2$ has a similar effect to increasing $r_2$; the mutation-selection balance point approaches the absorbing state, and the net-effect of selection away from the absorbing state is reduced, leading to a decrease in the fixation time. In line with the previous literature \cite{iwasa2004stochastic,haeno2013stochastic}, increasing the mutation rate increases the probability of tunneling.

In both panels of Fig. \ref{fig:fig6} the theoretical predictions from the WKB method are in excellent agreement with simulation results. This is the case even at the moderate population size of $N=100$.
Small deviations occur when mutation rates are low (dashed lines and open symbols in Fig.~\ref{fig:fig6}). The theory then slightly underestimates the fixation time (overestimates the fixation probability). This is a consequence of assuming that the population approaches the metastable state in a negligible amount of time. For very small mutation rates, it takes an increasing period of time for successful (i.e. non-vanishing) mutant lineages to appear. Deviations between theory and simulation results occur when $(1-u_2)r_1 \simeq r_2$. At this point the theory breaks down as the fixed point on the 1--2 edge approaches the absorbing state. The barrier associated with adverse selection is then negligible and the assumptions underlying the WKB-approximation are no longer justified.

\subsection{Region II}
To reach fixation in this Region the population must accumulate successive mutations of lower fitness ($r_0\gtrsim r_1 \gtrsim r_2$). The population first approaches a metastable state corresponding to the mutation-selection balance point of all three species. From here there are two possible routes to fixation, sequential or (almost) simultaneous extinction of types 0 and 1, as described previously and shown in Fig.~\ref{fig:fig3}. By computing the action accumulated along both routes, we have shown that the path of least action -- the most probable path to fixation -- corresponds to the path of sequential extinction.
We treat this two-hit process as two separate problems: (i) Escape from the interior metastable state to the boundary (loss of the advantageous type-0 phenotype); (ii) Escape from the boundary metastable state to the absorbing all-2 state (analogous to Region I). A typical realization of this sequence of events is shown in Fig.~\ref{fig:fig3}C.

As in Region I, the probability of tunneling decreases as the fitness advantage of type-1 cells over type-2 cells increases. This is because the fixed point on the 1--2 edge approaches the $n_1=N$ state. For the same reason, the tunneling rate decreases as the mutation rates decrease.

Following the convention used by \citeN{gottesman2012multiple}, we labeled the time to reach the 1--2 boundary as $\tau_{32}$, indicating that the 3-species system turns into a 2-species system when the wild-type cell goes extinct. Similarly, the time to travel from the boundary fixed point (2 species present in the population) to the absorbing state (1 species) is denoted by $\tau_{21}$. With this notation we also labeled the action accumulated along each segment as $S_{32}$ and $S_{21}$. As $\tau_{21}$ is given by Eq.~(\ref{eq:fixationtime}), we express the mean fixation time of type-2 cells as 
\begin{equation}
\tau = \frac{C_{32}}{\sqrt{N}} e^{N S_{32}} +\tau_{21}. \label{eq:taufitII}
\end{equation}
The coefficient $C_{32}$ is found by fitting to simulation data for the time taken to reach the 1--2 boundary as a function of the population size.

\begin{figure}
\begin{center}
\includegraphics*[width=0.5\textwidth]{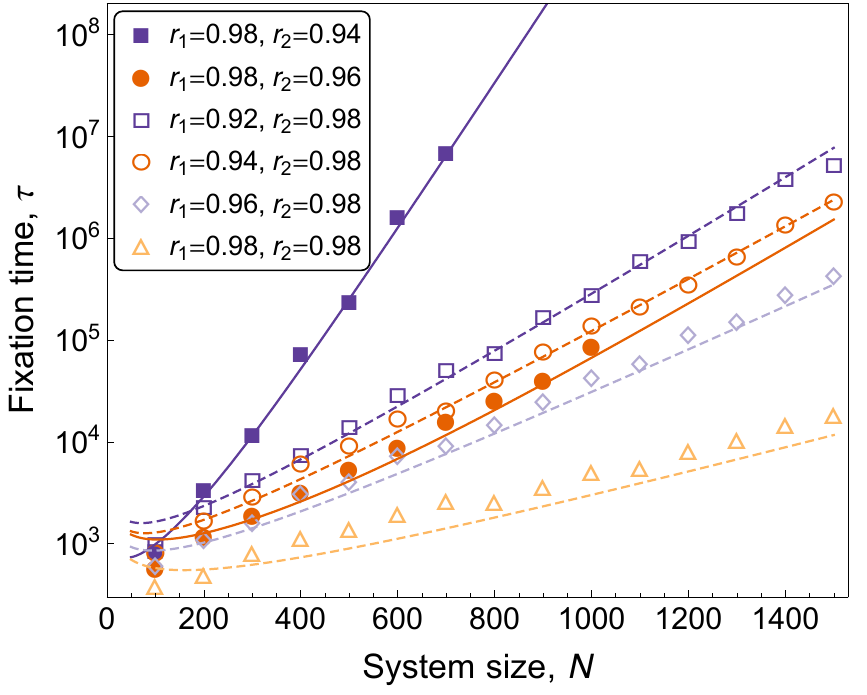}
\end{center}
\caption{\textbf{Results for Regions II and III.}
Fixation time in Region II (filled symbols) and Region III (empty symbols) of Fig.~\ref{fig:fig2} as a function of system size from simulations, averaged over $100$ realizations. Lines are from the theory, see Eq.~(\ref{eq:taufitII}) for Region II (solid lines) and Eq.~(\ref{eq:taufit}) for Region III (dashed lines). Remaining parameters are $r_0=1.0$ and $u_1=u_2=10^{-2}$.}
\label{fig:fig7}
\end{figure}

Small changes to the parameters now have significant effects on the fixation time, as shown in Fig.~\ref{fig:fig7} (filled symbols/solid lines). Increasing the fitness of the type-2 cells moves both the interior fixed point and the boundary fixed point towards the absorbing all-2 state. It also reduces the strength of selection away from the absorbing state. These combined effects dramatically reduce the mean fixation time, and its rate of increase with the population size.

\subsection{Region III}
In this Region fixation is controlled by the escape from the interior metastable state; it is a one-hit process.
The stable interior fixed point, corresponding to the mutation-selection balance point of the three species, is located close to the all-wild-type state as type-0 cells are the most advantageous. The direct path from the metastable state to the all-2 state is the dominant (least action) path, as shown in Fig.~\ref{fig:fig3}B. As a result the probability of tunneling is higher than in the previous cases. It increases as the fitness of type-2 cells and the mutation rates increase as the stable interior fixed point moves to lower numbers of type-1 cells (away from the all-1 state).

The fixation time is computed from Eq.~(\ref{eq:taufit}), where $S=S_{31}$ is the action accumulated along the direct path from the stable interior fixed point to the all-2 state. The coefficient $C_{31}$ is again found by fitting to simulation data. We see in Fig.~\ref{fig:fig7} (empty symbols/dashed lines) that varying the model parameters has a lesser effect on fixation times than in Region II. In Region III, fixation is a one-hit process -- the population only has to escape the stable fixed point -- and not a two-hit process as in Region II where the effects of the two steps are compounded. Contrary to the results for Region I, the mean fixation time is a decreasing function of $r_1$ in Region III. This can be explained as follows: by increasing $r_1$, the selection strength away from the 1--2 boundary decreases and the stable state moves to higher type-1 numbers, such that the population has an improved chance of reaching the 1--2 boundary. From there selection is directed towards the absorbing state, and the time spent on the 1--2 boundary is negligible compared to the time to reach this edge. Hence, the fixation time reduces as type-1 cells become more fit. The rate of increase of the fixation time with the population size reduces as well (see Fig.~\ref{fig:fig7}).

Note that there are systematic deviations between theory and simulation results in the data set shown as open triangles in Fig.~\ref{fig:fig7}, and to a lesser extent also for the data shown as open diamonds. This is attributed to the fact that the fitness parameters $r_1$ and $r_2$ are very similar to each other or equal for these instances, and they are also close to the fitness of the wild-type. Selection is then close neutral and the metastable state is only weakly attracting. The WKB approach then reaches its limits as the assumption of a long-lived metastable state begins to break down.

\section{Discussion}
In this paper we investigated the fixation of two successive mutations in a finite population of individuals proliferating according to the Moran process. We discussed this in the context of the somatic evolution of a compartment of cells. The accumulation of two mutations can correspond to the inactivation of a tumor suppressor gene or alteration of genes causing chromosomal instability (CIN) \cite{weinberg}. If the cell carrying two mutations is deleterious, as can be the case with recessive CIN genes \cite{michor2005can}, it will generally have low concentrations within the tissue. Then the chance of a cancerous phenotype emerging (further mutation) is very low. Demographic fluctuations can drive the double mutant to higher numbers, but these states are short-lived. If the double-mutant reaches fixation, the state is maintained until a further mutation occurs and hence the chance of a cancerous phenotype emerging is much greater.

We first analyzed the deterministic limit of the evolutionary dynamics. We identified parameter regimes in which mutation and selection balance. These are regimes in which the double mutant is not the most advantageous in the sequence. In finite populations, this mutation-selection balance gives rise to long-lived metastable states. Our analysis identified the escape from these metastable states as the key bottleneck to fixation of cells with two mutations. For parameter values for which there is no mutation-selection balance (i.e. type-2 cells have the highest fitness), the fixation dynamics is largely governed by the deterministic flow. The rate-limiting steps are then the appearance of successful mutant lineages \cite{nowak2004evolutionary}, and the subsequent fixation of cells with two mutations is a zero-hit process. As such the progression from healthy tissue (all wild type) to susceptible tissue (all type 2; inactivated TSG) will be fast relative to the cases in which a mutation-selection balance exists. If there is one stable fixed point in the deterministic dynamics, the process becomes a one-hit phenomenon limited by the escape from the corresponding metastable state. In regions with two fixed points one observes a two-hit process. The population becomes trapped in a first metastable state, escapes to a second metastable state, and then reaches full fixation.

In addition to this qualitative classification, we calculated fixation times in parameter regimes previously inaccessible to existing analytical approaches. These are precisely the regions of parameter space in which mutation-selection balance exists. We used the WKB-method to calculate the mean escape time from the corresponding metastable states to the absorbing all-type-2 state. This escape time is identified as the fixation time of type-2 cells. We tested the analytical expressions and numerical results obtained from the WKB approach against individual-based simulations of the population dynamics. Our theoretical predictions in principle rely on a limit of large but finite populations, and so they can be expected to be valid only for large enough populations. The comparison against simulations demonstrates the accuracy of our theory even at moderate population sizes of $N=100$ cells. For populations much smaller than this the assumptions of the WKB method break down. The rate-limiting step is then the occurrence of a successful lineage of mutants and not the escape from metastable states. The expressions obtained from the WKB approach become more accurate as the population size increases.

This analysis allowed us to classify how changes to the fitness landscape, mutation rates, and population size affect the probability of tunneling and the time-to-fixation of cells harboring two mutations. In terms of the development of tumors, our analysis shows that the path to accumulating mutations is not simply limited by the mutation rates, but also by escape from metastable states. Populations can exist in a heterogeneous state for very long periods of time before fluctuations eventually drive the second mutation to fixation. The probability with which stochastic tunneling occurs is, in part, determined by the location of these metastable states. If they are located close to the all-type-1 state, then the probability of tunneling is low. This occurs when cells with one mutation have a higher fitness than those with two mutations (Regions I and II). The probability of tunneling decreases as the fitness gap between these two types of cell increases or as mutation rates decrease. The mean fixation time increases exponentially as the fitness gap increases. Cells with one and two mutations are present in the tissue compartment for long periods of time; their numbers are maintained by the mutation-selection balance. Cell types are lost sequentially. Wild-type cells can be driven to extinction by selection (Region I) or by demographic fluctuations (Region II). The extinction of type-1 cells is driven exclusively by fluctuations. When type-2 cells have a higher fitness than type-1 cells, and when both are less fit than the wild-type (Region III), selection is always against cells of type-1. Mutation-selection balance maintains a low concentration of type-1 cells in the tissue, and hence the probability of tunneling is high. In this regime the mean fixation time is a decreasing function of the fitness of type-1 cells. As for all escape problems from metastable states, the fixation time scales exponentially with the size of the population. Fixation is noise-driven, and as the population size is increased the noise strength decreases, and hence fixation takes longer.

Although our theory is aimed at large population sizes and exponentially growing fixation times, we have shown that it can also make accurate predictions on biologically relevant timescales. Assuming a cell generation lasts for one day, our theory can capture fixation times of around 3 years or more ($>10^3$ generations). Related studies on the progression of cancer suggest a typical timescale on the order of $10$ years to accumulate a sufficient number of mutations \cite{beerenwinkel2007genetic,bozic2010accumulation}, which is well within the scope of our theory. However, the times predicted by our theory are extremely sensitive to parameter variation. This limits the parameter ranges for which biologically relevant timescales can be generated. Specifically selective (dis)advantages need to be small ($\lesssim10\%$). This is in agreement with selection coefficients in related studies \cite{bozic2010accumulation}. Of course the length of a cellular generation can vary by an order of magnitude or so, depending on the specific cell type \cite{weinberg}.

Our results do, however, allow an extrapolation to situations when fixation times become very long, for instance for very large populations and/or when selection is strongly against the invading mutants. In these scenarios, stochastic simulations can become too expensive computationally to provide meaningful measurements. Analytical methods based on backward master equations or backward Fokker-Planck equations suffer from computational limitations as well in such cases. Our mathematical work complements existing analytical approaches to the Moran model of cells acquiring two successive mutations. Previous work has provided an appropriate machinery with which to compute the time-to-fixation of the second mutation in situations without metastable states. The present paper specifically addresses cases in which fixation is limited by the escape from long-lived states. This contribution closes a gap in the analytical characterization of fixation in this model and a more complete picture is now available. We have added a new method to the toolbox used to study stochastic tunneling. Our deterministic analysis provides a systematic procedure to determine which tool to use for a given set of parameters. This accomplishment removes the need for stochastic simulations altogether, or at the very least it limits the circumstances under which they are needed.

The present work has clear limitations in that it focuses on the Moran model with only two successive mutations. We have not considered any processes beyond the second mutation, however such cases can exist in physical systems. If the type-2 cells are not cancerous, one would be interested in, for example, calculating when a metastatic cell (three mutations) first arises. In general this does not require the fixation of type-2 cells, and is related to the total number of type-2 cells over time. If metastable states are present, the cumulative number of type-2 cells prior to fixation is small, as described above. While we do not analyze this further, the typical number of type-2 cells at any time can in principle be computed from the quasi-stationary distribution, Eq.~(\ref{eq:quasistationaryapprox}).

Our systematic approach, along with the combined theoretical apparatus of previous work and the WKB method are readily transferable to more complex models of cancer initiation and progression.  One possible extension to this study is the generalization to more than two mutations. If a cell can accumulate $d$ possible mutations, metastable states are found provided $r_d<\max_{i=0,1,\dots,d-1}r_i$. A similar analysis can then be carried out. If the fitness landscape is arranged such that $r_0>r_1>\dots>r_d$, the problem is analogous to Muller's ratchet \cite{muller1964relation}, which describes the accumulation of successive maladaptive mutations. \citeN{metzger2013distribution} recently studied a special case of this problem using a WKB approach. Finally \citeN{gokhale2009pace} studied valley crossing dynamics with $d$ possible mutations. They have shown that allowing multiple paths to accumulate the mutations reduces the fixation time. As such allowing multiple paths in our model could reduce the fixation times we have measured. Work along both of these lines is in progress.

\section{Acknowledgments}
P.A. acknowledges support from the Engineering and Physical Sciences Council (UK), EPSRC. F.M. acknowledges support from the Dana-Farber Cancer Institute Physical Sciences-Oncology Center (NIH U54CA143798). T.G. acknowledges support from the EPSRC, grant reference EP/K037145/1.
The published article is available at \href{http://www.genetics.org}{www.genetics.org}.

%\newpage
\bigskip

\appendix

\makeatletter
\def\@seccntformat#1{\appendixname\ \csname the#1\endcsname:\quad}
\makeatother

\section{Fixed points and stability}\label{app:FP}
From the deterministic equations (\ref{eq:determinsticequations}) it can be seen that the state $\vec{x}=(0,0,1)$ is a fixed point, i.e. $\dot{x}_i=0$ at this point for $i=0,1,2$. This is the absorbing state, so this result is rather obvious. Non-trivial fixed points exist away from the absorbing state in some parameter regions.

The stability of a fixed point is determined by the eigenvalues of the Jacobian of the deterministic equations~(\ref{eq:determinsticequations}). Due to the overall constraint $x_0+x_1+x_2=1$, the system is effectively two-dimensional. We can write the Jacobian in terms of two variables, $x_1$ and $x_2$, as
\begin{equation}
\mathcal{J}(x_1,x_2)=
\begin{pmatrix}
\partial_{x_1} \dot{x}_1 & \partial_{x_2} \dot{x}_1 \\
\partial_{x_1} \dot{x}_2 & \partial_{x_2} \dot{x}_2
\end{pmatrix}. \label{eq:jacobian}
\end{equation}

Along the 1--2 boundary of the concentration simplex, Eqs. (\ref{eq:determinsticequations}) can be expressed in terms of a single variable, $x_1$. A fixed point, $x_1^*$, on this boundary satisfies the equations
\begin{eqnarray}
\dot{x}_1 &=& \bigl[(1-u_2)r_1-\rbar\bigr]x_1^*=0 \nonumber\\
\dot{x}_2 &=& u_2r_1x_1^*+(r_2-\rbar)(1-x_1^*)=0,
\end{eqnarray}
where $\rbar=r_1x_1^*+r_2(1-x_1^*)$. These equalities are satisfied by the value $x_1^*=1-u_2r_1/(r_1-r_2)$ (along with $x_2^*=1-x_1^*$). The parameter range in which this fixed point exists is determined by the condition $0<x_1^*<1$, which we can write as $r_2<(1-u_2)r_1$. The fixed point on the 1--2 edge therefore exists when type-1 cells have a fitness advantage over type-2 cells, the factor $1-u_2$ accounts for effects of mutation. Increasing this fitness advantage moves the fixed point towards $x_1=1$, or equivalently away from the absorbing state at $x_1=0$. For vanishing mutation rate $u_2$, the fixed point approaches the $x_1=1$ state.

Evaluating the eigenvalues of the $2\times 2$ Jacobian in Eq. (\ref{eq:jacobian}) at this fixed point, we find that the point is stable if $(1-u_2)r_1>(1-u_1)r_0$, and that it is a saddle if $(1-u_2)r_1<(1-u_1)r_0$. These two cases correspond to Regions I and II in Fig. \ref{fig:fig2}.

\medskip

A fixed point of Eqs.~(\ref{eq:determinsticequations}) with $x_0>0$ is found as
\begin{eqnarray}
x_1^* &=& \frac{[(1-u_1)r_0-r_2]u_1r_0}{u_2r_1(r_0-r_2)+(r_0-r_1)[(1-u_1)r_0-r_2]}, \nonumber\\
x_2^* &=& \frac{u_1u_2r_0r_1}{u_2r_1(r_0-r_2)+(r_0-r_1)[(1-u_1)r_0-r_2]}, \label{eq:fp}
\end{eqnarray}
provided the model parameters satisfy $(1-u_1)r_0>(1-u_2)r_1$ and $(1-u_1)r_0>r_2$. Further analysis of the Jacobian (\ref{eq:jacobian}) at this point shows that the fixed point is stable whenever it exists. This is the region of parameter space in which cells with one and two mutations respectively are both less fit than the wild-type. This is the case in Regions II and III in Fig. \ref{fig:fig2}. The fixed point moves closer to $x_0=1$ when the fitness advantage of the wild type cells is increased (e.g. by lowering the fitness of type-1 cells, $r_1 \to 0$). Decreasing the mutation rates also moves the fixed point closer to $x_0=1$.

%\newpage
\section{Solving the quasi-stationary master equation}\label{app:sol}
In terms of the variable $\vec{x}=\vec{n}/N$, we can write the QSME (\ref{eq:QSME}) for $\vec{n}\ne\vec{n}_{abs}$ as
\begin{equation}
0\approx -\frac{1}{\tau}\psi(\vec{x}) =
\sum_{\vec{v}} \left[ Nw_{\vec{v}}\left(\vec{x}-\frac{\vec{v}}{N}\right) \psi\left(\vec{x}-\frac{\vec{v}}{N}\right) - Nw_{\vec{v}}(\vec{x}) \psi(\vec{x}) \right],
\label{eq:QSMEx}
\end{equation}
where $w_{\vec{v}}(\vec{x})=T^{\vec{v}}_{N\vec{x}}/N$ and $\psi(\vec{x})=\pi_{N\vec{x}}$. Substituting the WKB ansatz, Eq. (\ref{eq:WKBansatz}), into Eq. (\ref{eq:QSMEx}) and expanding in powers of $N^{-1}$ we arrive at
\begin{eqnarray}
0&=&\sum_{\vec{v}} w_{\vec{v}}(\vec{x}) \left\{ e^{\vec{v}\cdot\vec{\nabla}S(\vec{x})} \left[ 1+\frac{1}{N}\vec{v}\cdot\vec{\nabla}S_1(\vec{x})-\frac{1}{2}\frac{(\vec{v}\cdot\vec{\nabla})^2}{N}S(\vec{x}) \right] -1 \right\}\nonumber\\
&&-\sum_{\vec{v}} e^{\vec{v}\cdot\vec{\nabla}S(\vec{x})} \frac{\vec{v}\cdot\vec{\nabla}}{N}w_{\vec{v}}(\vec{x})
+\mathcal{O}(N^{-2}), \label{eq:QSMEexpanded}
\end{eqnarray}
where we have ignored the term $\mathcal{O}\bigl((N\tau)^{-1}\bigr)$ as this term is smaller than $\mathcal{O}(N^{-2})$ ($\tau$ scales as $e^N$).
The leading-order terms of this equation are equivalent to the Hamilton-Jacobi equation,
\begin{equation}
H(\vec{q},\vec{p})=\sum_{\vec{v}} w_{\vec{v}}(\vec{q})\left(e^{\vec{v}\cdot\vec{p}}-1\right)=0, \label{eq:HJ}
\end{equation}
where $\vec{q}=\vec{x}$ is the so-called `position' variable, and $\vec{p}=\vec{\nabla}_{\vec{q}}S(\vec{q})$ is the so-called `momentum' variable. This equation is best solved using the method of characteristics, i.e. we look for parametric solutions, $(\vec{q},\vec{p})=(\vec{q}(t),\vec{p}(t))$. These trajectories fulfill Hamilton's equations,
\begin{equation}
\dot{\vec{q}} = \vec{\nabla}_{\vec{p}}H(\vec{q},\vec{p}), ~~ \dot{\vec{p}} = -\vec{\nabla}_{\vec{q}}H(\vec{q},\vec{p}). \label{eq:ham}
\end{equation}
They satisfy the principle of least action, and correspond to the most likely path taken in the so-called phase-space, the space spanned by $(\vec{q},\vec{p})$.
The Hamilton-Jacobi equation has the trivial solution $\vec{p}=\vec{0}$, which corresponds to the deterministic `relaxation' trajectory, for which the equation of motion is simply
\begin{equation}
\dot{\vec{q}}=\left.\vec{\nabla}_{\vec{p}}H(\vec{q},\vec{p})\right|_{\vec{p}=\vec{0}}=\sum_{\vec{v}} \vec{v}w_{\vec{v}}(\vec{q}).
\end{equation}

As we are interested in escape from a stable fixed point, we seek the non-trivial `activation' trajectory, for which $\vec{p}_a(\vec{q}) \ne \vec{0}$ in general. The relevant boundary condition is $\vec{p}_a(\vec{q}^{\;*})=\vec{0}$, where $\vec{q}^{\;*}$ indicates the fixed point of the deterministic dynamics from which the trajectory emanates.

In one-dimension, i.e. in the case of a single fixed point on the 1--2 boundary (Region I of Fig.~\ref{fig:fig2}), the Hamilton-Jacobi equation~(\ref{eq:HJ}) can be written as
\begin{equation}
H(q,p) = w_+(q)\left(e^p-1\right)+w_-(q)\left(e^{-p}-1\right)=0,
\end{equation}
where $q=x_1$ is the concentration of cells of type 1, $p= S'(q)$, $w_+(q)=T^{2\to1}(q)/N$, and $w_-(q)=T^{1\to2}(q)/N$ (reaction rates as described in Eq.~(\ref{eq:transitionrates}) ). This equation can be solved to obtain the activation trajectory $p_a(q)$, and hence $S(q)$.
We can now substitute $p_a(q)$ into the equation consisting of next-leading-order terms ($\mathcal{O}(N^{-1})$) of Eq.~(\ref{eq:QSMEexpanded}) to find $S_1(q)$.
Following this procedure we find
\begin{equation}
S(q)=-\int dq\; \ln\left[\frac{w_+(q)}{w_-(q)}\right], ~~ S_1(q)=\frac{1}{2}\ln\left[w_+(q)w_-(q)\right]. \label{eq:SS1}
\end{equation}
The QSD is now determined up to a normalization factor. The QSD is peaked about the fixed point located at $q^*=x_1^*=1-u_2r_1/(r_1-r_2)$, see Appendix~\ref{app:FP}. Hence we can expand the QSD~(\ref{eq:WKBansatz}) about this fixed point such that
\begin{equation}
\psi(q) \approx C\exp\left\{ -N\left[S(q^*)+\frac{1}{2} (q-q^*)^2 S''(q^*)\right]-S_1(q^*)-\dots\right\}, \label{eq:WKBsol}
\end{equation}
where we have used $S'(q^*)=p_a(q^*)=0$. Normalizing to unity then determines the normalization coefficient,
\begin{equation}
C= \sqrt{\frac{S''(q^*)}{2\pi N}}e^{NS(q^*)+S_1(q^*)}. \label{eq:norm}
\end{equation}

The QSD determined above breaks down when $n_1 \lesssim N^{1/2}$, or equivalently when $q \lesssim N^{-1/2}$, i.e. close to the absorbing state. In this region we consider a recursive solution of Eq.~(\ref{eq:QSME}) that does not rely on a specific form for the QSD, i.e. we do not use the WKB ansatz~(\ref{eq:WKBansatz}). We expand Eq.~(\ref{eq:QSME}) about $n_1=0$ to obtain
\begin{equation}
0=\sum_{v=\pm1} w'_v(0)\left[(n_1-v)\pi_{n_1-v}-n_1\pi_{n_1} \right].
\end{equation} 
This is to be solved for $\pi_{n_1}$ ($1<n_1\lesssim N^{1/2}$). Using $w'_-(0)=1$, we can write this as
\begin{equation}
f_{n_1+1}=\bigl[1+w'_+(0)\bigr]f_{n_1}-w'_+(0)f_{n_1-1},
\end{equation}
where $f_{n_1}=n_1\pi_{n_1}$. This recursive system can be solved to arrive at
\begin{equation}
\pi_{n_1}=\frac{\pi_1 \{1-[w'_+(0)]^{n_1}\} }{n_1[1-w'_+(0)]} \simeq \frac{\pi_1 [w'_+(0)]^{n_1} }{n_1[w'_+(0)-1]}, \label{eq:recursive}
\end{equation}
where the second step follows from $w'_+(0)>1$.
Using Eq.~(\ref{eq:timescaleansatz}) and expanding the relevant transition rate, $T_{\vec{n}_{\rm abs}-\vec{v}}^{\vec{v}} = T_{(0,1,N-1)}^{1 \to 2}$, about $n_1=0$ we can write $\pi_1\simeq\frac{1}{\tau}$. By matching the recursively obtained boundary-layer solution of Eq.~(\ref{eq:recursive}) with the WKB solution in Eq.~(\ref{eq:WKBsol}) at $0<q\lesssim N^{-1/2}$, we obtain an expression for the fixation time $\tau$, as shown in Eq.~(\ref{eq:fixationtime}).

%\newpage
\section{Numerical solutions for fixation time}\label{app:num}
We now address the case in which there is an internal stable fixed point of the deterministic dynamics. The problem then retains two degrees of freedom. We follow the initial steps of Appendix~\ref{app:sol} to arrive at the Hamilton-Jacobi equation~(\ref{eq:HJ}). Given that the original system is two-dimensional we now find four variables for the Hamilton-Jacobi problem, two position variables $q_1$ and $q_2$ (equivalent to $x_1$ and $x_2$), and two corresponding momenta $p_1$ and $p_2$. These are defined by $\vec{p}=\vec{\nabla}_{\vec{q}}S(\vec{q})$. As the `energy' is fixed ($H=0$) we have three effective degrees-of-freedom and no obvious solution to $H(\vec{q},\vec{p})=0$. We consider again Hamilton's equations (\ref{eq:ham}).
These equations describe the trajectory that minimizes the action, and hence by solving these we can then determine the fixation time.
To determine the boundary conditions we need to find the fixed points of Eqs.~(\ref{eq:ham}). We first note that there are three zero-momentum fixed points, which correspond to the fixed points of the deterministic equations (\ref{eq:determinsticequations}). Following~\citeN{gottesman2012multiple}, we label these as $M_1$ for the absorbing state ($\vec{q}=(0,0,1)$), $M_2$ for the 1--2 boundary fixed point, and $M_3$ for the stable interior fixed point defined by Eq.~(\ref{eq:fp}). As we seek to determine the activation trajectory, we need to find fixed points of Eqs.~(\ref{eq:ham}) with non-zero momenta, but with positions corresponding to $M_1$ and $M_2$ (the possible end points of the trajectories). These so-called `fluctuational fixed points' are labeled as $F_1$ and $F_2$.

The relevant trajectory is then found using an iterative method to solve the two-boundary problem. Consider the scenario in which the stable interior fixed point $M_3$ is the only fixed point of the deterministic system for $x_1$ and $x_2$, other than the absorbing state, i.e. for parameters in Region III of Fig.~\ref{fig:fig2}. Here the activation trajectory that leads to fixation starts at $M_3$ and finishes at $F_1$. To start the iteration, we fix the momenta for all times to the values at $F_1$, and then numerically integrate the equations of motion~(\ref{eq:ham}) for the position vector $\vec{q}$ forward in time, starting at $M_3$ and keeping the momenta constant. This integration is carried out for a sufficient range of time to reach the vicinity of the fixed point $F_1$, but not too long to avoid numerical errors building up. In the next step the relations for the momenta in Eq.~(\ref{eq:ham}) are then integrated backward in time using the trajectory $\vec{q}(t)$ found in the previous iteration. The momenta at the start of this backward integration are chosen as those corresponding to $F_1$. This procedure is then iterated, with alternating forward and backward integration of Hamilton's equation. At each step of the procedure the action of the path is found as
\begin{equation}
S_{31}=\int_{0}^{t} dt'\; \vec{p}(t')\cdot \dot{\vec{q}}(t').
\end{equation}
The iteration of alternating forward and backward integration is then repeated until $S_{31}$ has reached convergence. The action can then be used in Eq.~(\ref{eq:taufit}) to determine the fixation time $\tau_{31}$.

In Region II of parameter space a similar procedure is applied. In this case the minimizing trajectory in $(\vec{p},\vec{q})$-space which connects $M_3$ and $F_2$.

\end{document}